\documentclass[preprint,aps,prd,showpacs,preprintnumbers]{revtex4-1}

\usepackage{graphicx}
\usepackage{xcolor}
\usepackage{color}
\usepackage{amsmath,amsfonts,amssymb}
\usepackage{latexsym}
\usepackage[utf8]{inputenc}
\begin{document}
\title{Is a pole type singularity an alternative to inflation?}

\author{Marek Szyd{\l}owski}
\email{marek.szydlowski@uj.edu.pl}
\affiliation{Astronomical Observatory, Jagiellonian University, Orla 171, 30-244 Krakow, Poland}
\affiliation{Mark Kac Complex Systems Research Centre, Jagiellonian University, {\L}ojasiewicza 11, 30-348 Krak{\'o}w, Poland}
\author{Aleksander Stachowski}
\email{aleksander.stachowski@doctoral.uj.edu.pl}
\affiliation{Astronomical Observatory, Jagiellonian University, Orla 171, 30-244 Krakow, Poland}

\begin{abstract}
In this paper, we apply a method of reducing the dynamics of FRW cosmological models with the barotropic form of the equation of state to the dynamical system of the Newtonian type to detect the finite scale factor singularities and the finite-time singularities. In this approach all information concerning the dynamics of the system is contained in a diagram of the potential function $V(a)$ of the scale factor. Singularities of the finite scale factor manifest by poles of the potential function. In our approach the different types of singularities are represented by critical exponents in the power-law approximation of the potential. The classification can be given in terms of these exponents. We have found that the pole singularity can mimick an inflation epoch. We demonstrate that the cosmological singularities can be investigated in terms of the critical exponents of the potential function of the cosmological dynamical systems. We assume the general form of the model contains matter and some kind of dark energy which is parameterized by the potential. We distinguish singularities (by ansatz about the Lagrangian) of the pole type with the inflation and demonstrate that such a singularity can appear in the past. 
\end{abstract}

\maketitle

\section{Introduction}

The future singularity seems to be fundamental importance in the context of the observation acceleration phase of the expansion of the current universe. While the astronomical observations support the standard cosmological model $\Lambda$CDM we are still looking for the nature of dark energy and dark matter. In the context of an explanation of an acceleration conundrum it appears different theoretical ideas of a substantial form of dark energy as well as modification of the gravity \cite{Nojiri:2009pf}. For cosmological models with a different form of dark energy it is possible to define some form of the effective equation of state $p_\text{eff}=p_\text{eff}(\rho_\text{eff})$, where $\rho_\text{eff}$ is the effective energy density. For such a model coefficient equation of state $w_\text{eff}=\frac{p_\text{eff}}{\rho_\text{eff}}$, which is very close to value of $-1$ corresponding to the cosmological constant. In consequence in the future evolution of the universe can appear some new types of singularities. It was discovered by Nojiri et al. \cite{Nojiri:2005sx} that phantom/quintessence models of dark energy, for which $w_\text{eff}\simeq -1$, may lead to one of four different finite time future singularities. Our understanding of the finite scale factor singularity is the following. The singularities at which $a_\text{a}$ assumes the finite value, we called the finite scale factor. An appearance of future singularities is a consequence of a violation of the energy condition and may arise in cosmologies with phantom scalar fields, models with interaction of dark matter with dark energy as well as modified gravity theories \cite{Jimenez:2016sgs, Fernandez-Jambrina:2014sga}.

All types of finite late-time singularities can be classified into five categories following divergences of the cosmological characteristics \cite{Nojiri:2005sx, Fernandez-Jambrina:2014sga}:
\begin{itemize}
\item Type I (Big-Rip singularity): As $t\rightarrow t_\text{s}$ (finite), scale factor diverges, $a\rightarrow\infty$, and energy density as well as pressure also diverges, $\rho\rightarrow\infty,$ $|p|\rightarrow\infty$. They are classified as strong \cite{Caldwell:2003vq, FernandezJambrina:2006hj}.
\item Type II (Typical sudden singularity): As $t\rightarrow t_\text{s}$ (finite), $a\rightarrow a_\text{s}$ (finite), $\rho\rightarrow \rho_\text{s},$ $|p|\rightarrow\infty$. Geodesics are not incomplete in this case \cite{Barrow:2004xh, Barrow:2004hk, Barrow:1986}.
\item Type III (Big freeze): As $t\rightarrow t_\text{s}$ ,$a\rightarrow a_\text{s}$, and $\rho$ diverges, $\rho\rightarrow \infty$, as well as $|p|\rightarrow\infty$. In this case there is no geodesic incompleteness and they can be classified as weak or strong \cite{BouhmadiLopez:2006fu}.
\item Type IV (Generalized sudden singularity). As $t\rightarrow t_\text{s}$, $a\rightarrow a_\text{s}$ (finite value), $\rho\rightarrow \rho_\text{s},$ $|p|\rightarrow p_\text{s}$. Higher derivatives of the Hubble function are diverged. These singularities are weak \cite{Dabrowski:2013sea}.
\item Type V ($w$ singularities): As $t\rightarrow t_\text{s}$, $a\rightarrow \infty$, $\rho\rightarrow \infty$, $|p|\rightarrow 0$ and $w=\frac{p}{\rho}$ diverges. These singularities are weak \cite{Dabrowski:2009kg, Dabrowski:2009pc, FernandezJambrina:2007sx}.
\end{itemize}

It is interesting that singularities of type III appear in vector-tensor theories of gravity \cite{Jimenez:2008au, Jimenez:2009py}, while singularities of type II can appear in the context of a novel class of vector field theories basing on generalized Weyl geometries \cite{Jimenez:2016opp}.

The problem of obtaining constraints on cosmological future singularities from astronomical observations was investigated for all five types of singularities: type I in \cite{Keresztes:2014jua}, type II in \cite{Denkiewicz:2012bz}, type III in \cite{Balcerzak:2012ae}, type IV in \cite{Denkiewicz:2011uz}, type V in \cite{Jimenez:2009py}.

In this paper, we propose complementary studies of future singularities in the framework of cosmological dynamical systems of the Newtonian type. For the FRW cosmological models with the fluid, which are described by the effective equation of state $p_\text{eff}=p_\text{eff}(\rho_\text{eff})$ and $\rho_\text{eff}=\rho_\text{eff}(a)$, the dynamics of the model, without lost of degree of generality, can be reduced to the motion of a particle in the potential $V=V(a)$ \cite{Mortsell:2016too}. In this approach, a fictitious particle mimics of the evolution of a universe and the potential function is a single function of the scale factor, which reconstructs its global dynamics.

Our methodology of searching for singularities of the finite scale factor is similar to Odintsov et al. \cite{Nojiri:2005sx, Dabrowski:2009kg, Balcerzak:2012ae, Nojiri:2004ip, Kohli:2015yga, Perivolaropoulos:2016nhp} method of detection of singularities by postulating the non-analytical part in a contribution to the Hubble function. In our approach we assume that singularities are related with the lack of the analyticity in the potential itself or its derivatives. Additionally we postulate that in the neighborhood of the singularity, the potential as a function of the scale factor mimics the behavior of the poles of the function.
The advantage of our method is connected strictly with the additive non-analytical contribution to the potential with energy density of fluids which is caused by lack of analyticity of the scale factor or its time derivatives. This contribution arises from dark energy or dark matter.

In the paper, we also search pole types singularities in FRW cosmology models in the pole inflation model. These types of singularities are manifested by the pole in the kinetic part of the Lagrangian. In this approach in searching for singularities, we take ansatz on the Lagrangian.

The aim of the paper is twofold. Firstly (section II) we consider future singularities in the framework of the potential function. Secondly (section III) we consider singularities in the pole inflation approach. In section IV we summarize our results.

\section{Future singularities in the framework of potential of dynamical systems of Newtonian type}

\subsection{FRW models as dynamical system of Newtonian type}

We consider homogeneous and isotropic universe with a spatially flat space-time metric of the form
\begin{equation}\label{frw2}
ds^2=dt^2-a^2(t)\left[dr^2+r^2(d\theta^2+\sin^2\theta d\phi^2)\right],
\end{equation}
where $a(t)$ is the scale factor and $t$ is the cosmological time.

For the perfect fluid, from the Einstein equations, we have the following formulas for $\rho(t)$ and $p(t)$
\begin{align}
\rho &= 3H^2=\frac{3\dot{a}^2}{a^2},\label{hubble} \\
p &= -\frac{2\ddot{a}}{a}-\frac{\dot{a}^2}{a^2},\label{pressure}
\end{align}
where $\dot{ }\equiv\frac{d}{dt}$, $H\equiv\frac{\dot{a}}{a}$ is the Hubble function.

We assume that $\rho(t)=\rho(a(t))$ and $p(t)=p(a(t))$ depend on the cosmic time through the scale factor $a(t)$. From equations (\ref{hubble}) and (\ref{pressure}) we get the conservation equation in the form
\begin{equation}
\dot{\rho}=-3H(\rho+p).\label{cons}
\end{equation}

Equation (\ref{hubble}) can be rewritten in the equivalent form
\begin{equation}
\dot{a}^2=-2V(a),\label{friedmann}
\end{equation}
where
\begin{equation}
V(a)=-\frac{\rho_\text{eff}(a)a^2}{6} \label{potential_V}
\end{equation}
where $\rho_{\text{eff}}(a)$ is the effective energy density. For the standard cosmological model potential $V(a)$ is given by
\begin{equation}
V(a)=-\frac{\rho_{\text{eff}} a^2}{6}=-\frac{a^2}{6}\left(\rho_{m,0}a^{-3}+\rho_{\Lambda,0}\right),
\end{equation}
where $\rho_{\text{eff}}=\rho_{\text{m}}+\Lambda$ and $\rho_{\text{m}}=\rho_{\text{m,0}}a^{-3}$.
From equation (\ref{hubble}) and (\ref{pressure}), we can obtain the acceleration equation in the form
\begin{equation}
\frac{\ddot{a}}{a}=-\frac{1}{6}(\rho+3p),
\end{equation}
The equivalent form of the above equation is
\begin{equation}
\ddot{a}=-\frac{\partial V}{\partial a}.\label{pot}
\end{equation}

Due to equation (\ref{pot}), we can interpret the evolution of a universe, in dual picture, as a motion of a fictitious particle of unit mass in the potential $V(a)$. The scale factor $a(t)$ plays the role of a positional variable. Equation of motion (\ref{pot}) has the form analogous to the Newtonian equation of motion.

From the form of effective energy density, we can find the form of $V(a)$. The potential $V(a)$ determines the whole dynamics in the phase space $(a,\: \dot{a})$. In this case, the Friedmann equation (\ref{friedmann}) is the first integral and determines the phase space curves representing the evolutionary paths of the cosmological models. The diagram of potential $V(a)$ has all the information which are needed to construct a phase space portrait. Here, the phase space is two-dimensional
\begin{equation}
	\left\{ (a,\dot{a}) \colon \frac{\dot{a}^2}{2}+V(a)=-\frac{k}{2} \right\}
\end{equation}
and the dynamical system can be written in the following form
\begin{align}
\dot{a} &= x,\label{dota} \\
\dot{x} &= -\frac{\partial V(a)}{\partial a}.\label{dotx}
\end{align}

The lines $\frac{x^2}{2}+V(a)=-\frac{k}{2}$ represent possible evolutions of the universe for different initial conditions.

We can identify any cosmological model by the form of the potential $V(a)$. From the dynamical system (\ref{dota})-(\ref{dotx}) all critical points correspond to vanishing of right-hand sides of the dynamical system $\left(x_0=0, \left. \frac{\partial V(a)}{\partial a}\right|_{a=a_0}=0\right)$.

From the potential function $V(a)$, we can obtain the cosmological functions such as
\begin{equation}
t=\int^a \frac{da}{\sqrt{-2V(a)}},
\end{equation}
the Hubble function
\begin{equation}
H(a)=\pm\sqrt{-\frac{2V(a)}{a^2}},
\end{equation}
the deceleration parameter
\begin{equation}
q=-\frac{a\ddot{a}}{\dot{a}^2}=\frac{1}{2}\frac{d\ln(-V)}{d\ln a},
\end{equation}
the effective barotropic factor
\begin{equation}
w_\text{eff}(a(t))=\frac{p_\text{eff}}{\rho_{\text{eff}}}=-\frac{1}{3}\left(\frac{d\ln(-V)}{d\ln a}+1\right),
\end{equation}
the parameter of deviation from de Sitter universe \cite{FernandezJambrina:2006hj}
\begin{equation}
h(t)\equiv-(q(t)+1)=-\frac{1}{2}\frac{d\ln(-V)}{d\ln a}-1
\end{equation}
(note that if $V(a)=-\frac{\Lambda a^2}{6}$, $h(t)=0$), an effective matter density,
\begin{equation}
\rho_{\text{eff}}=-\frac{6 V(a)}{a^2},
\end{equation}
an effective pressure
\begin{equation}
p_{\text{eff}}=\frac{2 V(a)}{a^2}\left(\frac{d\ln(-V)}{d\ln a}+1\right),
\end{equation}
the first derivative of an effective pressure with respect of time
\begin{equation}
\dot{p}_{\text{eff}}=\frac{2\sqrt{-2V(a)}}{a}\left(\frac{d^2V(a)}{da^2}-\frac{2V(a)}{a^2}\right)
\end{equation}
and the Ricci scalar curvature (\ref{frw2})
\begin{equation}
R=\frac{6 V(a)}{a^2}\left(\frac{d\ln(-V)}{d\ln a}+2\right).
\end{equation}

\subsection{Singularities in terms of geometry of a potential function}

In this section we concentrate on two types of future singularities:
\begin{enumerate}
 \item finite-time singularities,
 \item finite scale factor singularities.
 \end{enumerate}
The finite time singularities can be detected using Osgood's criterion \cite{Osgood:1898}. We can simply translate this criterion to the language of cosmological dynamical systems of the Newtonian type. Goriely and Hyde formulated necessary and sufficient conditions for the existence of the finite time singularities in dynamical systems \cite{Goriely:2000}.

As an illustration of these methods used commonly in the context of integrability, it is considered a one-degree freedom Hamiltonian system with a polynomial potential. Such a system can be simply reduced to the form of the dynamical system of the Newtonian type \cite{Goriely:1998}. It is interesting that the analysis of the singularities of this system is straightforward when one considers the graph of the potential functions. These systems can possess a blow-up of the finite time singularities.

Following Osgood's criterion a solution $a(t)$ of the equation
\begin{equation}
\dot{a}=\sqrt{-2V(a(t))},
\end{equation}
with an initial problem
\begin{equation}
a(t_0)=a_0,
\end{equation}
blow up in the finite time if and only if
\begin{equation}
\int_{a_0}^\infty \frac{da}{\sqrt{-2V(a)}}<\infty.
\end{equation}

Let us assume that solutions becomes in a finite time $t_\text{s}$, for which $a=\phi(t)$ at $t=t_\text{s}$ diverges, $\phi(t_\text{s})=\infty$, where $t_\text{s}<\infty$. Then we have that
\begin{equation}
t_\text{s}=\int_{a_0}^\infty\frac{da}{\sqrt{-2V(a)}}+t_0<\infty.\label{integral}
\end{equation}
Moreover, a solution $a=\phi(t)$ is unique if
\begin{equation}
\int_{a_0}^{\phi(t)}\frac{da}{\sqrt{-2V(a)}}=\infty.
\end{equation}

If the potential assumes a power law $V=V_0 (a_\text{s}-a)^{\alpha}$, integral (\ref{integral}) does not diverge if only $\left(1-\frac{\alpha}{2}\right)$ is positive. In an opposite case, as $a\rightarrow a_\text{s}$, this integral diverges, which is an indicator of a singularity of a finite time $t\rightarrow t_\text{s}$ and $a\rightarrow\infty$.

In our further analysis we will postulate the form of the additive potential function $V(a)$ with respect to the effective energy density $\rho_\text{eff}$ (the interaction between the fluids is not considered)
\begin{equation}
V(a)=-\frac{\rho_\text{m}a^2}{6}+f(a)\equiv-\frac{\rho_\text{eff} a^2}{6},
\end{equation}
where $\rho_\text{m}=\rho_\text{m,0} a^{-3(1+w)}$ and $p_\text{m}=w\rho_\text{m}$, $w=const$ and choice of $f(a)$ is related with the assumed form of dark energy: $\rho_\text{de}=-\frac{6f(a)}{a^2}$.
Numerically one can simply detect these types of singularities. An analytical result can be obtained only for special choices of the function $f(a)$. In this context Chebyshev's theorem is especially useful \cite{Chen:2014fqa}. Following Chebyshev's theorem \cite{Tchebichef:1853, Marchisotto:1994} for rational numbers $p,\: q,\: r\: (r\neq 0)$ and nonzero real numbers $\alpha,\: \beta $, the integral
\begin{equation}
I=\int x^p \left(\alpha+\beta x^r\right)^q dx\label{int}
\end{equation}
is elementary if and only if at least one of the quantities $\frac{p+1}{r},\: q,\: \frac{p+1}{r}+q$ is an integer.

It is a consequence that integral (\ref{int}) may be rewritten as
\begin{equation}
\begin{split}
I &= \frac{1}{r}\alpha^{\frac{p+1}{r}+q}\beta^{-\frac{p+1}{r}}B_y\left(\frac{1+p}{r},\: q-1\right) \\ 
&= \frac{1}{p+1}\alpha^{\frac{p+1}{r}+q}\beta^{-\frac{p+1}{r}} y^{\frac{1+p}{r}} {_2} F_1\left(\frac{1+p}{r},\: 2-q,\: \frac{1+p+r}{r};\: y\right),\label{integral2}
\end{split}
\end{equation}
where $y=\frac{\beta}{\alpha}x^r$ and $B_y\left(\frac{1+p}{r},\: q-1\right)$ and ${_2}F_1\left(\frac{1+p}{r},\: 2-q,\: \frac{1+p+r}{r};\: y\right)$ are an incomplete beta function and a hyper-geometric function, respectively.

For the second distinguished singularity of a finite scale factor, the Chebyshev's theorem can be also very useful. In detection of these types of singularities, a popular methodology is to start from some ansatz on the function $a(t)$, which is near the singularity.

For example, let
\begin{equation}
a(t)=A+B\left(1-\frac{t}{t_\text{s}}\right)^n \Rightarrow B\left(1-\frac{t}{t_\text{s}}\right)^n =(a-a_\text{s}).
\end{equation}
Because $a(t=t_\text{s})=a_\text{s}$, $A=a_\text{s}$, the basic dynamical equation $\dot{a}^2=-2V(a)$ reduces to
\begin{equation}
\frac{d}{dt}(a-a_\text{s})=-\frac{nB}{t_\text{s}}\left(1-\frac{t}{t_\text{s}}\right)^{n-1}=-\frac{n B}{t_\text{s}}\left(\frac{a-a_\text{s}}{B}\right)^{\frac{n-1}{n}}
\end{equation}
Therefore
\begin{equation}
-2V(a)=-\frac{n}{t_\text{s}}B^{\frac{1}{n}}(a-a_\text{s})^{1-\frac{1}{n}},
\end{equation}
where $n<0$ i.e. $V(a)\propto (a_\text{s}-a)^\alpha$.

On the other hand if we postulate the above form of the potential one can integrate the equation of motion
\begin{equation}
\dot{a}^2=-2V(a)=V_0 (a_\text{s}-a)^\alpha,
\end{equation}
i.e.
\begin{equation}
a_\text{s}-a=\sqrt{V_0}\left(t_\text{s}-t\right)^{\frac{2-\alpha}{2}}.
\end{equation}
Therefore
\begin{equation}
a(t)=a_\text{s}+C(t_\text{s}-t)^n,\label{ansatz4}
\end{equation}
where $n=\frac{2-\alpha}{2}$.

This approach was considered in \cite{Denkiewicz:2011uz,Yurov:2017xjx}. We propose a similar approach, but we consider additionally the baryonic matter and the ansatz is defined by the potential $V(a)$.

In our approach to detection of future singularities it is more convenient ansatz for the form of the potential function rather than for directly for $a(t)$ function. We propose two ansatzes:
\begin{equation}
\dot{a}^2=\frac{1}{3}\rho_\text{m,0}a^{-1-3w}+V_0 (|a_\text{s}-a|)^\alpha,\label{ansatz2}
\end{equation}
\begin{equation}
\dot{x}^2=Bx^m+Cx^n,\label{ansatz}
\end{equation}
where $x=|a_\text{s}-a|$. For the second case, we neglect the matter effects. The implicit assumption (\ref{ansatz}) is that effects of matter are negligible near the singularity. For $a_\text{s}=0$, the above ansatzes are the same.

When we postulate a form of the potential, which is an additive function with respect different components of the fluid. We distinguish a part which arises from the barotropic matter and an additional part which gives the behavior of the potential in the neighborhood of poles (or its Pad{\'e} approximants). Our approach to the singularity investigation has its origin in Odintsov's paper \cite{Nojiri:2004ip}.

Let us integrate equation (\ref{ansatz}) with the help of Chebyshev's theorem
\begin{equation}
t(x)=\int^x\frac{dx}{\sqrt{Bx^m+Cx^n}}=\frac{1}{\sqrt{B}}\int^x x^{-\frac{m}{2}}\left(1+\frac{C}{B}x^{n-m}\right)^{-\frac{1}{2}}\,dx.
\end{equation}
Let us introduce new variable $u$
\begin{equation}
x^{-\frac{m}{2}}dx=du \quad \Rightarrow \quad x=\left(\frac{2-m}{2}u\right)^{\frac{2}{2-m}}.
\end{equation}
Then
\begin{equation}
t(x)=\frac{1}{\sqrt{B}}\int^x \left(1+\frac{C}{B}\left(\frac{2-m}{2}u\right)^{\frac{2(n-m)}{2-m}}\right)^{-\frac{1}{2}}\,du.\label{int2}
\end{equation}

For finding Chebyshev's first integral $I$ (eq. (\ref{integral2})) we check whether $\frac{p+1}{r}=\frac{2-m}{2(n-m)}$, $q=-\frac{1}{2}$, $\frac{p+1}{r}+q=\frac{2-n}{2(n-m)}$. For example for the general case if $m=-1-3w$ then $n=\frac{3(1+w)-2k(1+3w)}{2k}$, where $k\in \mathbb{Z}$. When the above conditions are kept then the solution of the (\ref{int2}) has the following form
\begin{equation}
t(a)-t_\text{s}=\frac{1}{\sqrt{B}}\frac{1}{1-\frac{m}{2}}|a_\text{s}-a|^{1-\frac{m}{2}} {_2} F_1\left(\frac{1-\frac{m}{2}}{n-m},\: 3/2,\: \frac{1-\frac{3m}{2}+n}{n-m};\: \frac{C}{B}|a_\text{s}-a|^{n-m}\right).\label{solutiontx}
\end{equation}

For the special case of (\ref{ansatz2}) for $a_\text{s}\approx 0$, the solution (\ref{solutiontx}) gives the following expression
\begin{equation}
t(a)-t_\text{s}=\frac{1}{\sqrt{\frac{1}{3}\rho_\text{m,0}}}\frac{1}{\frac{3+3w}{2}}a^{\frac{3+3w}{2}} {_2} F_1\left(\frac{\frac{3+3w}{2}}{\alpha+1+3w},\: 3/2,\: \frac{-\frac{1+3w}{2}+\alpha}{\alpha+1+3w};\: \frac{3V_0}{\rho_\text{m,0}}a^{\alpha+1+3w}\right).
\end{equation}

\subsection{Singularities for the potential $V=-\frac{1}{6}\rho_\text{m,0}a^{-1-3w}-\frac{V_0}{2} (|a_\text{s}-a|)^\alpha$}

\begin{figure}[ht]
	\centering
	\includegraphics[scale=1]{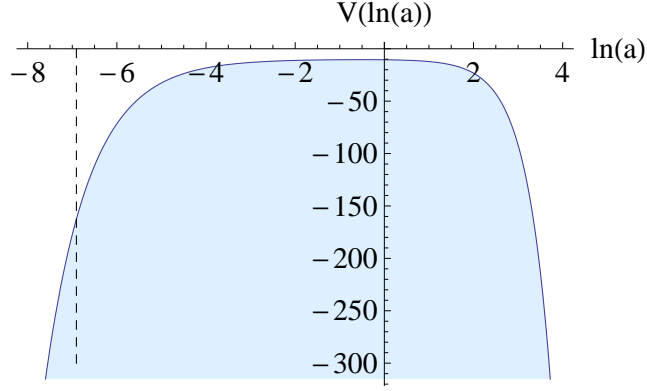}
	\caption{The diagram presents the potential function $V(a)$ (\ref{potential}) for $\alpha>1$ (generalized sudden singularity). The dashed line represents the generalized sudden singularities. This diagram corresponds with the phase portrait in Fig.~\ref{fig:2}.}
	\label{fig:4}
\end{figure}

\begin{figure}[ht]
	\centering
	\includegraphics[scale=1]{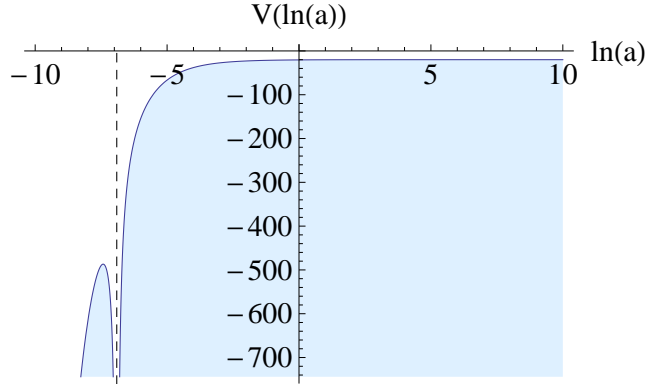}
	\caption{The diagram presents the potential function $V(a)$ (\ref{potential}) for $\alpha<0$ (big freeze). The dashed line represents the big freeze singularities.}
	\label{fig:9}
\end{figure}

\begin{figure}[ht]
	\centering
	\includegraphics[scale=1]{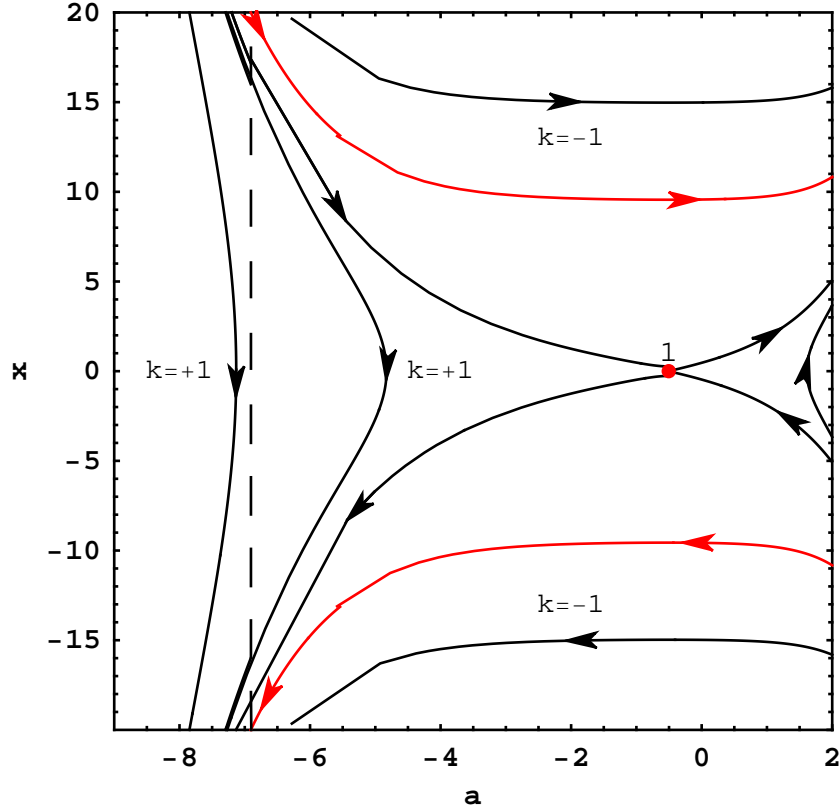}
	\caption{The diagram presents the phase portrait for dynamical systems (\ref{ds1})-(\ref{ds4}) for $w=1$ and $\alpha<2$. The dashed line represents the generalized sudden singularities. Critical point (1) presents the static Einstein universe and is the saddle type. The scale factor $a$ is presented in the logarithmic scale.}
	\label{fig:2}
\end{figure}

The potential $V(a)$ for ansatz (\ref{ansatz2}) is given by the following formula:
\begin{equation}
V=-\frac{1}{6}\rho_\text{m,0}a^{-1-3w}-\frac{V_0}{2} (|a_\text{s}-a|)^\alpha.\label{potential}
\end{equation}
Dabrowski et al. \cite{Dabrowski:2014fha} assumed that singularities can appear in the future history of the Universe. We consider singularities which are located in the future as well as in the past of the Universe.
The potential $V(a)$ for the best fit value (see section~III) is presented in Fig.~\ref{fig:4}. In this case, it appears the generalized sudden singularity. We show too the diagram of the potential when it appears the big freeze singularity (see Fig.~\ref{fig:9}). In this case, dynamical system (\ref{dota})-(\ref{dotx}) has the form
\begin{align}
\dot{a} &= x,\label{ds1} \\
\dot{x} &= -\frac{(1+3w)}{3}{\rho_\text{m,0}}a^{-2-3w}+\frac{\alpha}{2}V_0 \left(a_\text{s}-a\right)^{\alpha-1},\label{ds2}
\end{align}
for the $a_\text{s}>a$ and
\begin{align}
\dot{a} &= x,\label{ds3} \\
\dot{x} &= -\frac{(1+3w)}{3}{\rho_\text{m,0}}a^{-2-3w}-\frac{\alpha}{2}V_0 \left(a-a_\text{s}\right)^{\alpha-1},\label{ds4}
\end{align}
for the $a_\text{s}<a$. The phase portrait for the above dynamical system for the best fit value (see section III) is presented in Fig.~\ref{fig:2}.

Because the dynamical systems (\ref{ds1})-(\ref{ds2}) and (\ref{ds3})-(\ref{ds4}) are insufficient for introducing of the generalized sudden singularity, the above dynamical system can be replaced by three-dimensional dynamical system in the following form
\begin{align}
\dot{a} &=x ,\label{ds5} \\
\dot{x} &= \ddot{a}=y, \\
\dot{y} &= \dddot{a}=\frac{(1+3w)(2+3w)}{3}{\rho_\text{m,0}}a^{-1-3w}x(t)-\frac{\alpha(\alpha-1)}{2}V_0 \left(a_\text{s}-a\right)^{\alpha-2}x(t) \label{ds6}
\end{align}
for the $a_\text{s}>a$ and
\begin{align}
\dot{a} &= x,\label{ds7} \\
\dot{x} &= \ddot{a}=y, \\
\dot{y} &= \dddot{a}=\frac{(1+3w)(2+3w)}{3}{\rho_\text{m,0}}a^{-1-3w}x(t)-\frac{\alpha(\alpha-1)}{2}V_0 \left(a-a_\text{s}\right)^{\alpha-2}x(t) \label{ds8}
\end{align}
for the $a_\text{s}<a$.

\begin{figure}[ht]
	\centering
	\includegraphics[scale=1]{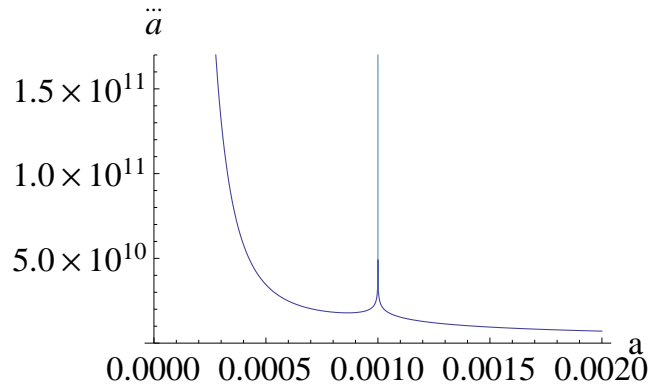}
	\caption{The diagram presents the evolution of $\dddot{a}$ with respect to the scale factor $a$ for the flat universe for $\alpha$ parameter equal 1.82. Note that, for $a=10^{-3}$, it appears the generalized sudden singularity.}
	\label{fig:7}
\end{figure}

\begin{figure}[ht]
	\centering
	\includegraphics[scale=1]{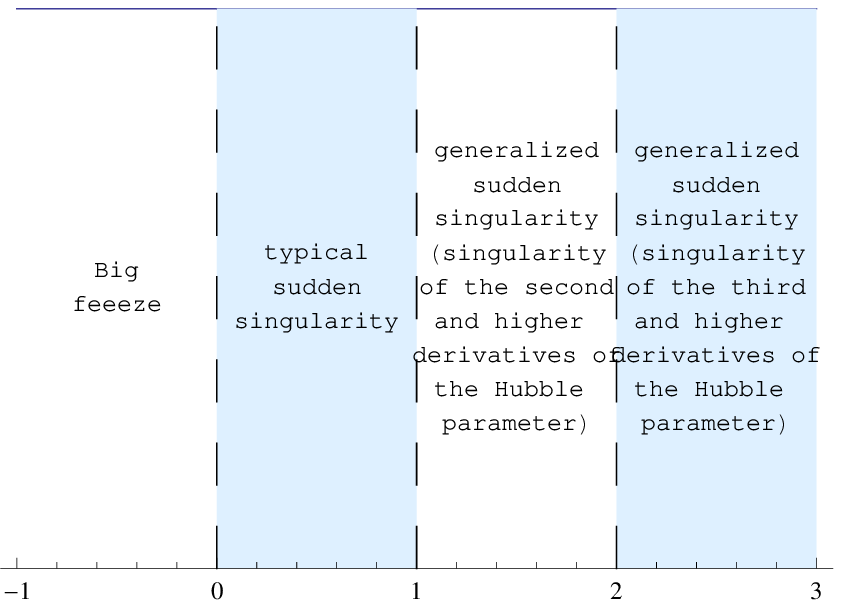}
	\caption{The diagram presents the generic types of singularity with respect to the value of the parameter $\alpha$.}
	\label{fig:5}
\end{figure}

If we investigate dynamics in terms of geometry of the potential function then their natural interpretation can be given. It means a lack of analyticity of the potential itself (in consequence $da/dt$ blows up) or its derivatives (higher order derivatives of the scale factor blow up). The singularities are hidden beyond the phase plane ($\dot{a}, a$) and we are looking for its in the enlarged phase space. In our approach to the detection of different types of the finite scale factor singularities we explore information contained in the geometry of the potential function, which determine all characteristics of singularities. This function plays an analogous role as the function $h(t)$ in the standard approach.

From the potential (\ref{potential}) we can obtain formula for the Hubble parameter
\begin{equation} \label{eq:53}
H(a)=\pm\sqrt{-2 V(a)}=\pm\sqrt{\frac{1}{3}\rho_\text{m,0}a^{-3-3w}+\frac{V_0}{a^2} (|a_\text{s}-a|)^\alpha}.
\end{equation}
Let us note $a_\text{s}\ll a$ we obtain the $\Lambda$CDM model. The first derivative of the Hubble function $\dot{H}$ is given by
\begin{equation}
\dot{H}=-\frac{1+w}{2}\rho_\text{m,0}a^{-3-3w}-\frac{\alpha V_0}{2 a} (a_\text{s}-a)^{\alpha-1}-\frac{ V_0}{ a^2} (a_\text{s}-a)^{\alpha},
\end{equation}
for $a_\text{s}>a$
and
\begin{equation}
\dot{H}=-\frac{1+w}{2}\rho_\text{m,0}a^{-3-3w}+\frac{\alpha V_0}{2 a} (a-a_\text{s})^{\alpha-1}-\frac{ V_0}{ a^2} (a-a_\text{s})^{\alpha},
\end{equation}
for $a_\text{s}<a$. Note that in the singularity $a=a_\text{s}$ if $\alpha<1$ then $\dot{H}=\pm\infty$.

The second derivatives of the Hubble function is
\begin{multline}
\ddot{H}=H[\frac{3(1+w)^2}{2}\rho_\text{m,0}a^{-3-3w}+ \frac{\alpha (\alpha-1)V_0}{2} (a_\text{s}-a)^{\alpha-2}\\ +\frac{\alpha V_0}{2 a^2} (a_\text{s}-a)^{\alpha-1}+\frac{\alpha V_0}{ a} (a_\text{s}-a)^{\alpha-1}+\frac{ 2V_0}{ a^2} (a_\text{s}-a)^{\alpha}],
\end{multline}
for $a_\text{s}>a$
and
\begin{multline}
\ddot{H}=H[\frac{3(1+w)^2}{2}\rho_\text{m,0}a^{-3-3w}+ \frac{\alpha (\alpha-1)V_0}{2} (a-a_\text{s})^{\alpha-2}\\ -\frac{\alpha V_0}{2 a^2} (a-a_\text{s})^{\alpha-1}-\frac{\alpha V_0}{ a} (a-a_\text{s})^{\alpha-1}+\frac{ 2V_0}{ a^2} (a-a_\text{s})^{\alpha}],
\end{multline}
for $a_\text{s}<a$.
Note that in the singularity $a=a_\text{s}$ if $\alpha<2$ then $\ddot{H}=\pm\infty$.

The effective matter density is given by
\begin{equation}
\rho_{\text{eff}}=-\frac{6 V(a)}{a^2}=\rho_\text{m,0}a^{-3-3w}+\frac{3V_0}{a^2} (|a_\text{s}-a|)^\alpha
\end{equation}
and the effective pressure has the following form
\begin{multline}
p_{\text{eff}}=\frac{2 V(a)}{a^2}\left(\frac{d\ln(-V)}{d\ln a}+1\right)=w\rho_\text{m,0}a^{-3-3w}+\alpha V_0 a^{-1} (a_\text{s}-a)^{\alpha-1}-V_0 a^{-2}(a_\text{s}-a)^\alpha
\end{multline}
for $a_\text{s}>a$ and
\begin{multline}
p_{\text{eff}}=\frac{2 V(a)}{a^2}\left(\frac{d\ln(-V)}{d\ln a}+1\right)=w\rho_\text{m,0}a^{-3-3w}-\alpha V_0 a^{-1} (a-a_\text{s})^{\alpha-1}-V_0 a^{-2}(a-a_\text{s})^\alpha
\end{multline}
for $a_\text{s}<a$. The first derivative of the effective pressure has the following form
\begin{equation}
\dot{p}_{\text{eff}}=\frac{1}{3\sqrt{2}a^3}V_0^{3/2}(a_\text{s}-a)^{-2+\frac{3\alpha}{2}}(2a_\text{s}^2-4a a_\text{s}+a^2(2+\alpha-\alpha^2))
\end{equation}
for $a_\text{s}>a$ and
\begin{equation}
\dot{p}_{\text{eff}}=\frac{1}{3\sqrt{2}a^3}V_0^{3/2}(a-a_\text{s})^{-2+\frac{3\alpha}{2}}(2a_\text{s}^2-4a a_\text{s}+a^2(2+\alpha-\alpha^2))
\end{equation}
for $a_\text{s}<a$.
Note that for the singularity $a=a_\text{s}$ if $\alpha<1$ then $p_{\text{eff}}=\pm\infty$ and if $\alpha>1$ then $p=0$.

The type of singularity with respect to the parameter $\alpha$ is presented in Fig.~\ref{fig:5}. 

For the classification purpose we take into account singularities located at a constant, non-zero value of the scale factor (we do not consider a singularity at $a=0$). This classification covers last five cases from Dabrowski's paper \cite{Dabrowski:2014fha}. Due to such a representation of singularities in terms of a critical exponent of the pole one can distinguish generic (typical) cases from non-generic ones. The classification of the finite scale factor singularities for the scale factor $a>0$  and the potential $V=-\frac{1}{6}\rho_\text{m,0}a^{-1-3w}-\frac{V_0}{2} (|a_\text{s}-a|)^\alpha$ \cite{Dabrowski:2014fha} is presented in Table \ref{table:2}. We call singularities as generic if the corresponding value of $\alpha$ parameter for such singularities is non zero measure. In the opposite case for $\alpha$ parameter assumes discreet value such singularities are fine-tuned.

It is interesting that in the case without matter, a $w$-singularity appears for the special choice of the parameter $\alpha$ ($\alpha=4/3$). Let us note that all singularities without the $w$-singularity are generic.

\begin{table}
	\caption{Values of critical parameter $\alpha$ for singularities with the finite scale factor \cite{Dabrowski:2014fha} for the scale factor $a>0$ and the potential $V=-\frac{1}{6}\rho_\text{m,0}a^{-1-3w}-\frac{V_0}{2} (|a_\text{s}-a|)^\alpha$.}
	\label{table:2}
	\begin{center}
		\begin{tabular}{ccccccc} \hline
			Name & $\rho(t_\text{s})$ & $p(t_\text{s})$ & $\dot{p}(t_\text{s})$ etc.& $w(t_\text{s})$ & $\alpha$& classification\\ \hline \hline
Type II & & & & & & \\ (Typical sudden singularity) & $\rho_\text{s}\neq 0$ & $\infty$	& $\infty$ & finite & $0<\alpha<1$ & weak			
 \\ \hline
Type II$_g$ & & & & & & \\ (Generalized sudden singularity) & $\rho_\text{s}\neq 0$ & $p_\text{s}\neq 0$	& $\infty$ & finite & $\alpha>1$ & weak						
 \\ \hline
Type III & & & & & & \\ (Big freeze) & $\infty$ & $\infty$ & $\infty$ & finite & $\alpha<0$ & weak or strong
 \\ \hline
Type IV & & & & & & \\ (Big separation) & 0 & 0 & $\infty$	& $\infty$ & $\alpha<4/3$ & weak \\
& & & & & ( for the case & \\
& & & & &  without matter) &
 \\ \hline
Type V & & & & & & \\ ($w$ singularity) & 0 & 0 & 0 & $\infty$ & $\alpha=4/3$ & weak \\
& & & & & ( for the case & \\
& & & & &  without matter) &
 \\ \hline
		\end{tabular}
	\end{center}
\end{table}

\subsection{Pad{\'e} approximant for the potential $V(a)$}

The standard methodolology of searching for singularities based on the Puiseux series \cite{FernandezJambrina:2010ev}. We proposed, instead the application of this series, using of the Pad{\'e} approximant for parametrization of the potential which has poles at the singularity point.

The second derivative of the nonanalitical part of the potential $V(a)$, which we note as $\ddot{\tilde V}(a)$, in the neighborhood of a singularity, can be approximated by a Pad{\'e} approximant. The Pad{\'e} approximant in order $(k,\:l)$, where $k>0$ and $l>0$ is defined by the following formula
\begin{equation}
P_{kl}(x) = \frac{c_0 + c_1 x + \cdots + c_l x^l}{1 + b_1 x + \cdots b_k x^k}.
\end{equation}

The coefficients of the Pad{\'e} approximant can be found by solving of the following system of equations
\begin{align}
P_{kl}(x_0) &= f(x_0) \\
P_{kl}'(x_0) &= f'(x_0) \\
& \vdots \nonumber \\
P_{kl}^{(k+l)}(x_0) &= f^{(k+l)}(x_0),
\end{align}
where $f$ is a function which is approximated.

Let $a>a_\text{s}$. For the potential $V(a)$ the Pad{\'e} approximant in order $(1,\:1)$ is given by
\begin{equation}
P_{11}(a) = \frac{\frac{2\ddot{\tilde{V}}(a_0)\tilde{V}^{(3)}(a_0)-2\tilde{V}^{(3)}(a_0)^2a_0+\ddot{\tilde{V}}(a_0)\tilde{V}^{(4)}(a_0)a_0}{2\tilde{V}^{(3)}(a_0)+\tilde{V}^{(4)}(a_0)a_0}+\frac{2\tilde{V}^{(3)}(a_0)^2-\ddot{\tilde{V}}(a_0)\tilde{V}^{(4)}(a_0)}{(2\tilde{V}^{(3)}(a_0)+\tilde{V}^{(4)}(a_0)a_0)}a}{1-\frac{\tilde{V}^{(4)}(a_0)}{(2\tilde{V}^{(3)}(a_0)+\tilde{V}^{(4)}(a_0)a_0)}a},
\end{equation}
where derivation is with respect to time, $a_0$ is the value of $a$ for which are calculated the coefficients of Pad{\'e} approximant. For $a>a_\text{s}$: $\ddot{\tilde{V}}(a)=-\frac{V_0 \alpha(\alpha-1)}{2}(a-a_\text{s})^{\alpha-2}$, $\ddot{\tilde{V}}(a)=-\frac{V_0 \alpha(\alpha-1)(\alpha-2)}{2}(a-a_\text{s})^{\alpha-3}$, $\ddot{\tilde{V}}(a)=-\frac{V_0 \alpha(\alpha-1)(\alpha-2)(\alpha-3)}{2}(a-a_\text{s})^{\alpha-4}$ and for $a<a_\text{s}$: $\ddot{\tilde{V}}(a)=-\frac{V_0 \alpha(\alpha-1)}{2}(a_\text{s}-a)^{\alpha-2}$, $\ddot{\tilde{V}}(a)=\frac{V_0 \alpha(\alpha-1)(\alpha-2)}{2}(a_\text{s}-a)^{\alpha-3}$, $\ddot{\tilde{V}}(a)=-\frac{V_0 \alpha(\alpha-1)(\alpha-2)(\alpha-3)}{2}(a_\text{s}-a)^{\alpha-4}$

In this case, for the Pad{\'e} approximant, a singularity appears when $a=a_\text{s}$.

\begin{figure}[ht]
	\centering
	\includegraphics[scale=1]{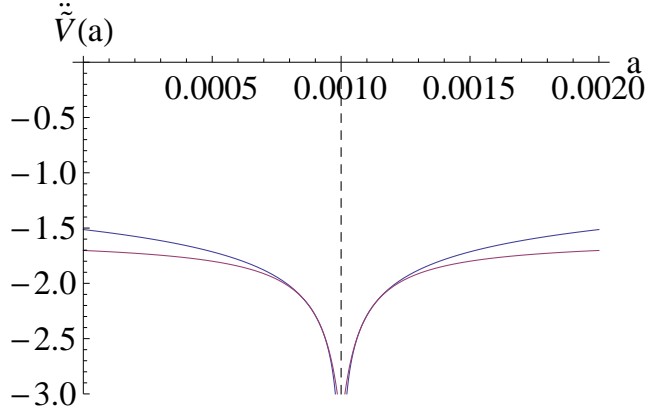}
	\caption{The diagram presents $\ddot{\tilde{V}}(a)$ (top curve) and an approximation of $\ddot{\tilde{V}}(a)$ by Pad{\'e} approximant (bottom curve) for the generalized sudden singularity. The dashed line presents the generalized sudden singularity.}
	\label{fig:6}
\end{figure}

In Fig. \ref{fig:6} is shown how Pad{\'e} approximant can approximate the potential $\ddot{\tilde V}(a)$ in the neighborhood of a singularity.

Pad{\'e} approximant is not only used to a better approximation of the behaviour of the potential or time derivatives near the singularity. It can be used directly in the basic formula $da/dt=\sqrt{-2V(a)}$ for defining nonregular parts of the potential. Therefore in our approach we can apply just this ansatz instead an ansatz for $a(t)$ like in the standard approach. On the background of Pad{\'e} exponents, we can make the following assumption
\begin{equation}
\dot{a}^2=-2V=-2V_\text{m}+P_{kl}=-\frac{\rho_{m,0} a^2}{6}+\frac{c_0 + c_1 a + \cdots + c_l a^l}{1 + b_1 a + \cdots b_k a^k}.
\end{equation}

\section{Singularities in the pole inflation}

Let us concentrate on pole types singularities in the FRW cosmology models. These types of singularities are manifested by the pole in the kinetic part of the Lagrangian. We also distinguish pole inflation singularities in following \cite{Saikawa:2017wkg,Galante:2014ifa,Terada:2016nqg}. In this approach in searching for singularities, we take an ansatz on the Langragian rather than scale factor postulated in the standard approach.

We consider dynamics of cosmological model reduced to the dynamical system of the Newtonian type, i.e, $\frac{d^2 a}{dt^2}= -\frac{dV}{da}$, where $a(t)$ is scale factor factor, $t$ is the cosmological time. Then the evolution of the Universe is mimicking by a motion of a particle of a unit mass in the potential which is a function of the scale factor only.

By pole singularities we understand such a value of the scale factor $a=a_\text{s}$ for which the potential itself jump to infinity or its $k$-order derivatives $(k=2,3,\ldots)$ with respect to the scale factor (in consequence we obtain jump discontinuities in the behaviour of the time derivatives of the scale factor).

In the pole inflation approach, beyond appearing of the inflation, the kinetic part of the Lagrangian has a pole or their derivatives have a pole. We use the pole inflation approach in the form which was defined in \cite{Saikawa:2017wkg,Galante:2014ifa,Terada:2016nqg}.

The Langragian has the following form \cite{Galante:2014ifa}
\begin{equation}
L=-3a \dot a^2-\frac{1}{2}a_\text{p}\rho^{-p}\dot\rho^2 a^3-V_0(1-c\rho)a^3,\label{langragian}
\end{equation}
where $a_\text{p}$, $p$, $V_0$ and $c$ are model parameters and $\dot{}\equiv \frac{d}{dt}$. 
Let $\rho=|a-a_\text{s}|^n$. Then Langragian (\ref{langragian}) can be rewritten as
\begin{equation}
L=-3a \dot a^2-\frac{n^2}{2}a_\text{p}|a-a_\text{s}|^{3n-p-2}\dot a^2 a^3-V_0(1-c|a-a_\text{s}|^n)a^3,\label{langragian2}
\end{equation}

After variation with respect to the scale factor $a$ we get the acceleration equation, which can be rewritten as
\begin{multline}
\ddot{a}(t)=\left[-3\dot a(t)^2
+\frac{1}{2}a(t)^2\text{\Large(}6V_0-2cV_0(a(t)-a_\text{s})^{n-1}((3+n) a(t)-3a_\text{s}) \right. \\
\left. -a_\text{p} n^2 (a(t)-a_\text{s})^{3n-p-3} ((1+3n-p)a(t)-3 a_\text{s}) 
\dot{a}(t)^2\text{\Large)} \right]
/\left[6a(t)+a_\text{p} n^2 a(t)^3 (a(t)-a_\text{s})^{3n-p-2}\right] \label{pole1}
\end{multline}
for $a(t)>a_\text{s}$ and
\begin{multline}
\ddot{a}(t)=\left[-3\dot a(t)^2
+\frac{1}{2}a(t)^2\text{\Large(}6V_0-2cV_0(a_\text{s}-a(t))^{n-1}(3a_\text{s}-(3+n) a(t)) \right. \\
\left. -a_\text{p} n^2 (a_\text{s}-a(t))^{3n-p-3} (3 a_\text{s}-(1+3n-p)a(t)) 
\dot{a}(t)^2\text{\Large)}\right]
/\left[6a(t)+a_\text{p} n^2 a(t)^3 (a_\text{s}-a(t))^{3n-p-2}\right] \label{pole2}
\end{multline}
for $a(t)<a_\text{s}$.

The first integral of (\ref{pole1}) and (\ref{pole2}) has the following form
\begin{equation}
H(t)=\frac{\sqrt{2V_0}\sqrt{1-c|a(t)-a_\text{s}|^n}}{\sqrt{6+a_\text{p}n^2a(t)^2|a(t)-a_\text{s}|^{3n-p-2}}}.\label{integral3}
\end{equation}

Let $a_\text{p}=-\frac{6c}{n^2}$ and $p=2n$. Then the first integral (\ref{integral3}) has the form
\begin{equation}
H(t)=\sqrt{\frac{V_0}{3}}\frac{\sqrt{1-c|a(t)-a_\text{s}|^n}}{\sqrt{1-ca(t)^2|a(t)-a_\text{s}|^{n-2}}},\label{integral4}
\end{equation}
which guarantees the inflation behaviour when $a_\text{s} / a(t) \ll 1$.

The slow roll parameters can be used to find the value of the model parameters. These parameters are defined as
\begin{equation}
\epsilon=-\frac{\dot H}{H} \text{ and } \eta=2\epsilon-\frac{\dot\epsilon}{2H\epsilon}.\label{slow1}
\end{equation}

The following relation exists between the scalar spectral index and the tensor-to-scalar ratio and the slow roll parameters
\begin{equation}
n_\text{s}-1=-6\epsilon+2\eta \text{ and } r=16\epsilon.\label{slow2}
\end{equation}

Let $a_\text{s}\ll a(t)$. If we use formulas (\ref{slow1}) and (\ref{slow2}), then we get equations for parameters $c$ and $a_\text{p}$
\begin{equation}
c=\frac{a_\text{fin}^{-n} r (16 (3 n - p) (1 + n - n_\text{s} - p) + 4 (-1 - 5 n + n_\text{s} + 2 p) r + 
   r^2)}{768 n^3 - 16 n^2 (40 p + r) + r (4 p + r) (4 (-1 + ns + p) + r) + 
 4 n (32 p^2 - 8 p r + (4 - 4 n_\text{s} - 3 r) r)},
\end{equation}
\begin{equation}
a_\text{p}=\frac{6 a_\text{fin}^{-3 n + p} r (4 (-1 + n + n_\text{s}) + r)}{n^2 (32 (6 n^2 - 5 n p + p^2) + 4 (-1 - 5 n + n_\text{s} + 2 p) r + r^2)},
\end{equation}
where $a_\text{fin}$ is the value of the scale factor in the end of the inflation epoch. Because we assume $a_\text{p}=-\frac{6c}{n^2}$ and $p=2n$, then we get that
\begin{equation}
c=a_\text{fin}^{-n}
\end{equation}
and the tensor-to-scalar ratio $r$ is given by
\begin{equation}
r=4(1-n_\text{s}).
\end{equation}
The best fit of the scalar spectral index $n_\text{s}$ is equal 0.9667 \cite{Ade:2015xua}. In consequence, $r=0.1332$.

Because in this model the singularity is in the begining of the inflation and we also assume that number of e-folds is equal 50, the value of $a_\text{s}$ is equal $e^{-50}a_\text{fin}\approx 1.93\times 10^{-22}a_\text{fin}$.

Up to now, the inflation has a methodological status of very interesting hypothesis added to the standard cosmological model. Note that in the context of pole singularities, the following question is open: Can pole singularities be treated as an alternative for inflation?

\begin{figure}[ht]
	\centering
	\includegraphics[scale=1]{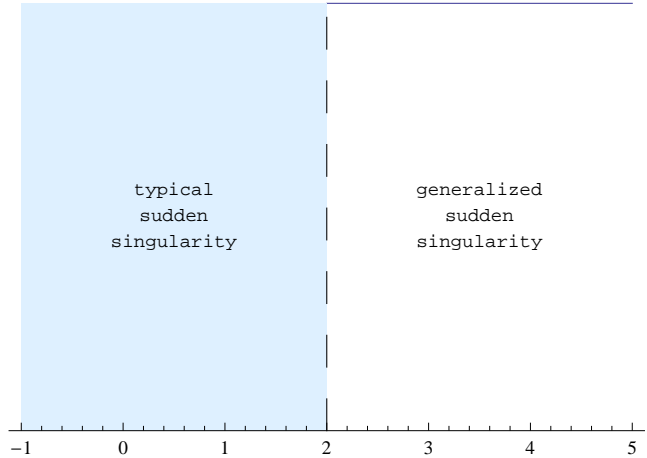}
	\caption{The diagram presents the generic types of singularity in the pole inflation model with respect to the value of parameter $n$.}
	\label{fig:8}
\end{figure}

The type of singularities in the model is depended on the value of $n$ parameter. If $n>2$ then the singularity in the model represents the generalized sudden singularity. The typical sudden singularity appears when $n<2$ (see Fig.~\ref{fig:8}).

\begin{figure}[ht]
	\centering
	\includegraphics[scale=1]{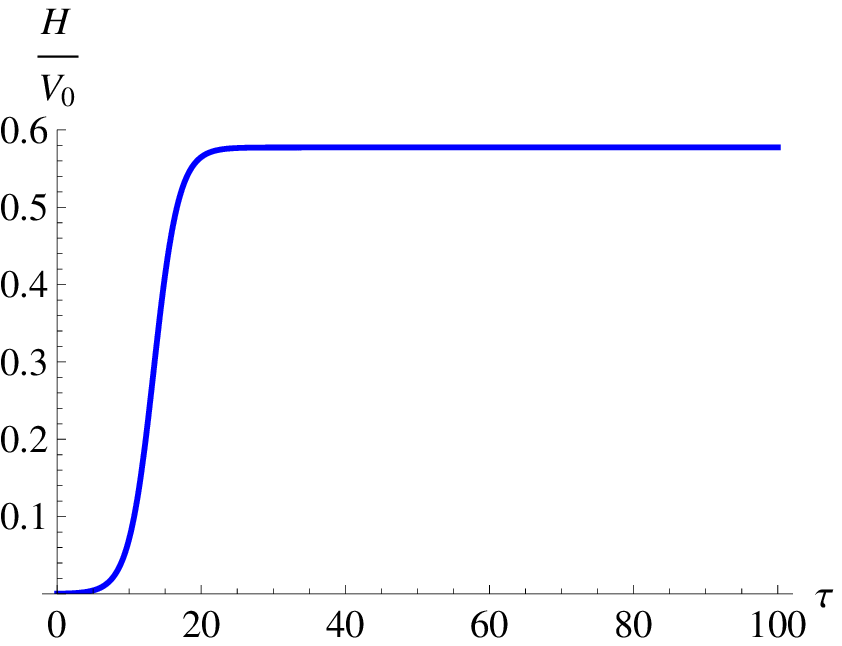}
	\caption{The diagram presents the evolution of $H/V_0$ in the pole inflation model for an example value $n=-1$. Dimensionless time $\tau$ is equal $V_0 t$. Here, $\tau=100$ corresponds the end of the inflation period ($t=10^{-32}\text{s}$).}
	\label{fig:1}
\end{figure}

\begin{figure}[ht]
	\centering
	\includegraphics[scale=1]{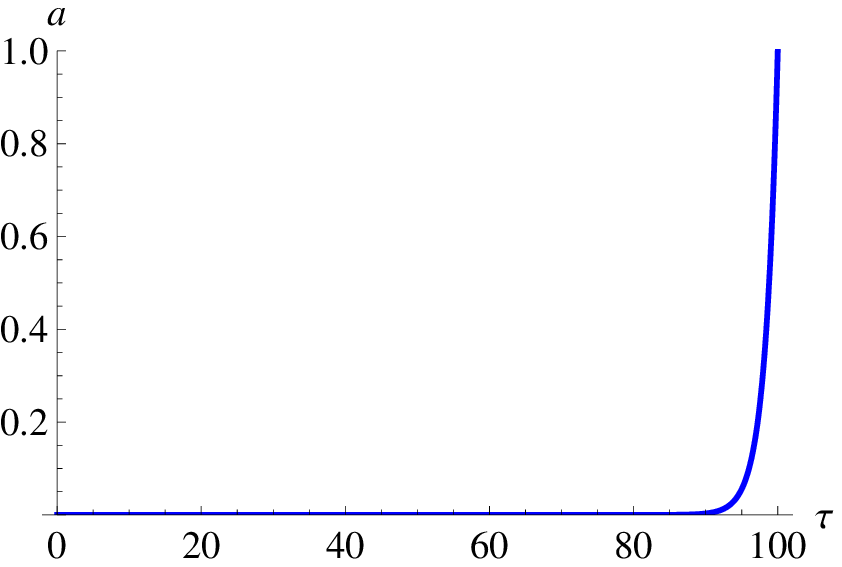}
	\caption{The diagram presents the evolution of the scale factor in the pole inflation model for an example value $n=-1$. Dimensionless time $\tau$ is equal $V_0 t$. Here, $\tau=100$ corresponds the end of the inflation period ($t=10^{-32}\text{s}$).}
	\label{fig:3}
\end{figure}

Figure~\ref{fig:1} presents the evolution of $H/V_0$ in the pole inflation model for an example value $n=-1$. Note that, in this case, typical sudden singularity appears. In this singularity, the value of the Hubble function is equal to zero. Figure~\ref{fig:3} presents the evolution of the scale factor in the pole inflation model for an example value $n=-1$.

We can use this approach to a description of the inflation in the past. This model also can be used to description of the behaviour of dark energy in the future. But it is possible only in the case when the generalized sudden singularity appears ($n>2$). In the case when the typical sudden singularity appears, the bounce appears in the singularity. In result, this case ($n<2$) cannot be considered as a model of the behaviour of dark energy in the future.

Our result is in agreement with the general statement that a physically reasonable cosmological models with the eternal inflation possess an initial singularity in the past \cite{Borde:1993xh,Borde:1996pt}. In the standard approach to the classification of singularities in the future, Borde and Vilenkin elided the fact that inflation in the past history takes place. Of course, if we assumed that inflation epoch was happened in the past corresponding results obtained without this ansatz should be corrected.

\section{Conclusions}

In the paper, we study the finite scale factors using a method of reducing dynamics of FRW cosmological models to the particle moving in the potential as a function of the scale factor. In the model we assume that universe is filled by matter and dark energy in the general form which are characterized by the potential function. The singularities in the model appears due to a non-analytical contribution in the potential function. Near the singularity point the behavior of the potential is approximated by poles.

Using of the potential method we detected the scale factor singularities. In the detection of singularities of the finite scale factor we used the methodology similar to the detection of singularities of the finite time \cite{Nojiri:2005sx}. An advantage of this method is that the additional contribution to the potential is additive and is strictly related with the form of energy density of dark energy.

Using of the method of the potential function gives us a geometrical framework of the investigation of singularities. The dynamics is reduced to the planar dynamical system in the phase space $(a,da/dt)$. The system possesses the first integral energy like for particle moving in the potential $\frac{1}{2}\left(\frac{da}{dt}\right)^2 +V(a)=E=const$. The form of the potential uniquely characterizes the model under consideration.

We demonstrated that finite scale factor singularities can be investigated in terms of critical exponent $\alpha$ of the approximation of the potential near the singularity point $a=a_\text{s} \colon V=V_0(a-a_\text{s})^{\alpha}$. The classification of singularities can be given according the value of $\alpha$ parameter. For the class of singularities under consideration the effects of visible matter near singularities are negligible in comparison to effects of dark energy modelled by the nonregular potential.

For better approximation the behavior of the potential near the singularities we apply the method of Pad{\'e} approximants.

In the general the behaviour of the system is approximated by the behaviour of the potential function near the poles. The singularities appear as the consequence a lack of analyticity of the potential or its derivatives with respect the scale factor. In consequence, the time derivatives of the scale factor with respect the cosmological time blows up to infinity.

For the generalized sudden singularity under consideration $da/dt, d^2a/dt^2$ as well us the Hubble parameter are regular and third derivative with respect to time blow up. Of course this type singularity cannot be visualised in the phase space $(a,da/dt)$ because higher dimensional derivatives are non-regular. Therefore we construct a higher dimensional dynamical system in which nonregular behaviour of $d^3 a/dt^3$ can be presented. Finally the dynamical system in which one can see this type of singularity has dimension three.

Our general conclusion is that framework of the particle like reducing cosmological dynamical sytems can be useful in the contex of singularities in FRW cosmology with barotropic form of equation of state. Different types of singularities have different and universal values of exponenents in a potential approximation near the singularities. We believe that this simple approach reveals a more fundamental connection of the singularity problem with an important area in physics -- critical exponents in complex systems.

It is interesting that the generalized sudden singularity is a generic feature property of modified gravity cosmology \cite{Barrow:2004hk} as well as brane cosmological models \cite{Shtanov:2002ek}.

Our method has an heuristic power which helps us to generalize some types of singularities. The advantage of the proposed method of singularities detection seems to be its simplicity. Our ansatz is rather for potential of cosmological system than for the scale factor. In order the potential function is an additive function of matter contribution $\rho_{\text{eff}}$ in oposition to the scale factor.

Let us consider a $w$-singularities case discovered by Dabrowski and Denkiewicz \cite{Dabrowski:2009kg}. After simple calculations one can check that the potential of the form $V= (a-a_s)^{4/3}$ admits generalized w-singularities when both $\rho_{\text{eff}}$ and $H$ are zero, $dp/dt$ goes to zero and $w$ diverges. Let us note that in the case of non-zero cosmological constant this type of singularity disappears automatically.

It was proposed to constrain the position of singularities basing directly on the ansatz on an approximation for the scale factor near the singularity \cite{Dabrowski:2009kg, Balcerzak:2012ae, Nojiri:2004ip, Kohli:2015yga, Perivolaropoulos:2016nhp}. It is model independent approach as it is based only on mathematics of singularity analysis. Then this scale factor approximation is used in cosmologicals models to determine a type of singularities and estimate model parameters. Alternative approach which we believe is methodologically proper is to consider a cosmological model and prove the existence of singularities in it. Of course, such singularities are model dependent. Then we estimate the redshift corresponding singularity and determine a type of the singulatity. This approach has been recently applied by Alam et al. \cite{Alam:2016wpf}. In their paper the position of possible future singularities is taken directly from the brane model and after constraining the model parameters one can calculate numerical value of singularity redshift. Note that, in the brane model, the generalized sudden singularity can appear in the future history of the universe \cite{Shtanov:2002ek,Antoniadis:2017rgz,Alam:2005pb}. For these singularities the potential function jump discontinuously following the corresponding pole singulariety.

In the standard approach of probing of singularities, it is considered the ansatz for prescribing of the asymptotic form of the scale factor $a(t)$. In our investigation, we search some special types of pole singularities called pole inflation singularities. In the study of appearing of these types of singularities, we make the ansatz by the Langragian of the model. This Langragian contains regular part as well as jump discontinuities. The jump discontinuities can appear in the kinetic part of the Lagrangian. Our estimation of slow roll parameters that existence pole inflation in the past history of the universe.

In our paper, we demonstrated that inclusion the hypothesis of the inflation in the past evolution of the Universe can modify our conclusions about their appearance and position during cosmic evolution. In the standard practice, the information about the inflation in the past is not included in the postulate for a prescribed asymptotic form of the scale factor $a(t)$. The situation can be analogical like in Vilenkin \cite{Borde:1993xh,Borde:1996pt} when the eternal inflation determines singularity of the big bang in the past.

\acknowledgments{We thank dr Adam Krawiec and dr Orest Hrycyna for insightful discussions.}


\end{document}